\documentclass[unsortedaddress,reprint]{revtex4-1}

\usepackage[utf8]{inputenc}
\usepackage{amssymb,latexsym}
\usepackage{amsbsy} 
\usepackage{amsmath,amsthm}

\usepackage[shortlabels]{enumitem}
\usepackage{hyperref}

\usepackage{graphicx}
\usepackage{color}

\usepackage{tikz}
\usepackage{algorithmic}
\usepackage{wrapfig}

\theoremstyle{plain}

\theoremstyle{remark}

\theoremstyle{definition}

\newcommand \ket[1] {| {#1} \rangle}

\newcommand \argmin {\mathop\mathrm{arg\,min}}

\newcommand{\dua}[2]{\langle #1, #2 \rangle}

\renewcommand{\epsilon}{\varepsilon}
\renewcommand{\kappa}{\varkappa}
\renewcommand{\phi}{\varphi}

\newcommand{\ham}{\mathcal{H}}
\newcommand{\opA}{\mathcal{A}}
\newcommand{\opK}{\mathcal{K}}
\newcommand{\ene}{\mathcal{E}}
\newcommand{\VC}{\mathcal{V}} 
\newcommand{\KP}{\mathrm{KP}}

\newcommand{\cc}{\mathrm{CC}}
\newcommand{\full}{\mathrm{full}}

\newcommand{\VV}{\mathbb{V}}

\newcommand \commentout[1] {}

\newcommand{\rev}[1]{\textcolor{black}{#1}}

\begin{document}

\title{Homotopy continuation methods for coupled-cluster theory in quantum chemistry}

\author{Fabian M. Faulstich}
\email{f.m.faulstich@berkeley.edu}
\affiliation{
Department of Mathematics, University of California, Berkeley}

\author{Andre Laestadius 
	}
\email{andre.laestadius@oslomet.no}
\affiliation{
Department of Computer Science, Oslo Metropolitan University}
\affiliation{Hylleraas Centre for Quantum Molecular Systems, Department of Chemistry, University of Oslo}

\date{\today}

\begin{abstract}
Homotopy methods have proven to be a powerful tool for understanding the multitude of solutions provided by the coupled-cluster polynomial equations. 
This endeavor has been pioneered by quantum chemists that have undertaken both elaborate numerical as well as mathematical investigations. 
Recently, from the perspective of applied mathematics, new interest in these approaches has emerged using both topological degree theory and algebraically oriented tools. 
This article provides an overview of describing the latter development. 
\end{abstract}

\maketitle

\section{Introduction}

Coupled-cluster (CC) theory is a widely acclaimed, high-precision wavefunction approach that is used in quantum chemistry and is of great interest to both practitioners as well as theoreticians~\cite{bartlett2007coupled}. 
The origin of CC theory dates back to 1958 when Coester proposed \rev{to use} an exponential parametrization of the wave function~\cite{coester1958bound}.
This parametrization was independently derived by Hubbard~\cite{hubbard1957description} and Hugenholtz \cite{hugenholtz1957perturbation} in 1957 as an alternative to summing many-body perturbation theory (MBPT) contributions order by order.
A milestone of CC theory is the work by {\v{C}}{\'i}{\v{z}}ek from 1966~\cite{vcivzek1966correlation}.
In this work, {\v{C}}{\'i}{\v{z}}ek discussed the foundational concepts of second quantization (as applied to many-fermion systems), normal ordering, contractions, Wick’s theorem, normal-ordered Hamiltonians (which was a novelty at that time), Goldstone-style diagrammatic techniques, and the origin of the exponential wave function ansatz.
He moreover derived the connected cluster form of the Schr\"odinger equation and proposed a general recipe for how to produce the energy and amplitude equations through projections of the connected cluster form of the Schr\"odinger equation on the reference and excited determinants, which was illustrated using the CC doubles (CCD) approximation. 
This work also reports the very first CC computations, using full and linearized forms of CCD, for nitrogen (treated fully at the \emph{ab initio} level) and benzene (treated with a PPP model Hamiltonian). 
For a more detailed history of CC theory, several reviews have been written by, e.g., K\"ummel~\cite{kummel1991origins} and {\v{C}}{\'i}{\v{z}}ek~\cite{vcivzek1991origins}. 
Other articles that provide insight into the history and development of CC theory include those by Bartlett~\cite{bartlett2005theory}, Paldus~\cite{paldus2005beginnings}, Arponen~\cite{arponen1991independent}, and Bishop~\cite{bishop1991overview}. 

The CC equations are a set of nonlinear algebraic equations that are typically solved using (quasi) Newton-type methods~\cite{press2007numerical}. 
\rev{Each equation in the set corresponds to a projection on a specific excitation of the reference determinant, and the number of equations increases rapidly with the size of the system and allowed excitations} under consideration. 
Since the CC equations are a set of nonlinear algebraic equations, there are multiple roots to the CC equations.
\rev{The existence of multiple roots to the CC equations can present challenges for the practical implementation of the method.}
For example, the roots can be difficult to adequately converge to, and the convergence properties of the iterative solution methods can strongly depend on the employed initial guess. 
Furthermore, excited states can also be targeted with equation-of-motion (EOM)-CC \cite{Rowe1968,monkhorst1977calculation,Koch1990,StantonBartlett1993,Koch1994}, where an initial ``ground-state'' CC calculation is the starting point of a non-Hermitian diagonalization problem.

Although great progress has been made along the fundamental and mathematical study of the root structure of the CC equations, which include homotopy methods applied to the CC methodology, progress and wide-spread applications have been hampered by the high dimensionality and non-linearity of the CC equations, as well as the steep scaling of {\it algebro-computational} methods. 
The first study on this topic dates back to 1978, where Živković and Monkhorst investigated the singularities and multiple solutions of the single-reference CC equations \rev{revealing conditions for reality and the maximum multiplicity of solutions}~\cite{vzivkovic1978analytic}. 
In 1998, Kowalski \rev{and Jankowski} revived homotopy methods in connection with CC theory and used them to solve a CCD system~\cite{kowalski1998towards}. 
This was followed by a fruitful collaboration with Piecuch, \rev{who worked on multiple solutions to the single reference CC and state-universal MRCC equations~\cite{paldus1993application,piecuch1990coupled}. Piecuch and Kowalski} extended the application of homotopy methods to CC single and doubles-(CCSD)-, CC singles, doubles and triples-(CCSDT)-, and CC single, doubles, triples and quadruples-(CCSDTQ)-equations~\cite{piecuch2000search} for a 4-electron system described by a minimum basis set. 
They also introduced the formalism of $\beta$-nested equations and proved the {\it Fundamental Theorem of the $\beta$-NE Formalism}, which enabled them to explain the behavior of the curves connecting multiple solutions of the various CC polynomial systems, i.e., from CCSD to CCSDT, CCSDT to CCSDTQ, etc. 
In~\cite{kowalski2000complete2}, Piecuch and Kowalski used homotopy methods to determine all solutions of nonlinear state-universal multireference CCSD equations based on the Jeziorski-Monkhorst ansatz, proving two theorems that provided an explanation for the observed intruder solution problem. 
In a sequel work~\cite{kowalski2000complete}, they used homotopy methods to obtain all solutions of the generalized Bloch equation, which is nonlinear even in a CI parametrization.
\rev{Further articles utilizing homotopy methods arose in the late '90s where, amongst other things, symmetry breaking processes in non-linear CC formulation are explained~\cite{kowalski1998full,jankowski1999physical1,jankowski1999physical2,jankowski1999physical3,jankowski1999physical4}.}

In addition to examining the latest developments in homotopy techniques in CC theory, this article aims to present new findings that result from the application of homotopy approaches using both topological degree theory and algebraic geometry tools, from an applied mathematics perspective. 
These approaches are essential for expanding the scope of mathematical investigations beyond the ground state. 
While local analyses using strong monotonicity~\cite{schneider2009analysis,rohwedder2013continuous,rohwedder2013error,laestadius2017analysis,faulstich2018analysis,laestadius2019coupled,faulstich2019numerical,Kvaal-et-al-ECC-2020,faulstich2023s} have been useful since Schneider's seminal work~\cite{schneider2009analysis}, they only provide a limited perspective compared to the broader study of the CC equations using topological degree theory or algebraic geometry~\cite{csirik2023coupled, faulstich2022coupled}. 
By adopting these tools, we can gain a more comprehensive understanding of the mathematical structure of the CC equations and their solutions, providing new avenues for further explorations and refinements of the theory.

This article is organized as follows: In Section~\ref{sec:CC-theory} we briefly introduce the CC theory including concepts like the CC variety as well as the equivalence between a wavefunction being an eigenstate and the CC parametrization (standard- and EOM-CC). In Section~\ref{sec:h-cont} we recall the concept of homotopy continuation techniques, highlighting important edge-cases to consider. In Section~\ref{sec:RootStructure}, we elaborate on the root structure of polynomial systems and the challenges involved with using (quasi) Newton-type methods. In Section~\ref{sec:Homotopy1} we then present upper bounds on the number of roots of the CCSD equations based on B\'ezout's theorem and outline how to establish improved bounds using the Bernstein-Khovanskii-Kushnirenko theorem. In Section~\ref{sec:Homotopy2}, we discuss the mathematical existence of solution curves connecting CC roots at different truncation levels. We also present an energy error estimate valid for any approximate eigenstate of exponential (CC) form. Both these results have appeared previously in the mathematical literature~\cite{csirik2023coupled}. We conclude in Section~\ref{sec:conclusion} by summing up the main points as well as presenting a brief outlook on future work.

\section{Coupled-cluster theory}
\label{sec:CC-theory}

The underlying idea of coupled-cluster (CC) theory is the exponential parametrization of the targeted wavefunction $\ket{\Psi}$, i.e, for a given reference determinant $\ket{\Phi_0}$, we make the ansatz
\begin{equation}
\ket{\Psi} = e^{\hat T} \ket{\Phi_0},
\end{equation}
where $\hat T$ is the new unknown, called the {\it cluster operator}, which we will define shortly. 
Using this ansatz, we find 
\begin{equation}
\label{eq:SimHam}
\mathcal{H} \ket{\Psi} =  E \ket{\Psi}
~\Leftrightarrow~
e^{- \hat T} \mathcal{H} e^{\hat T} \ket{\Phi_0} =  E  \ket{\Phi_0},
\end{equation}
where $\mathcal{H}$ is the considered Hamiltonian. 
Projecting Eq.~\ref{eq:SimHam} yields 
\begin{equation} 
\label{eq:projections}
\left\lbrace
\begin{aligned}
E &= \langle \Phi_0  \mid e^{- \hat T} \mathcal{H} e^{\hat T}  \mid \Phi_0\rangle \\ 
0 &= \langle \Phi  \mid e^{- \hat T} \mathcal{H} e^{\hat T}  \mid \Phi_0\rangle,\quad \forall \,\, \ket{\Phi} \perp \ket{\Phi_0}
\end{aligned}
\right.
\end{equation}
of which the latter is used to compute the cluster operator.
The cluster operator is a linear combination of \rev{elementary particle-hole excitation operators} (vide infra), and we shall henceforth highlight the dependence of $\hat T$ to its expansion coefficients by writing $\hat T(\bf t)$.
\rev{Elementary particle-hole excitation operators} are merely compositions of projections onto a subset of single-particle basis elements (the occupied orbitals) and wedge products with another subset of single-particle basis elements (the virtual orbital). 
It is, therefore, highly convenient to label the \rev{elementary particle-hole excitation operators} using multi-indices $\mu$ that clarify the projections and wedge products involved (see, e.g.,~\cite{schneider2009analysis}). 
Employing this notation yields the following expression for the cluster operator
\begin{equation}
\label{cluster matrices}
\hat T({\bf t}) = \sum_{\mu } t_\mu \hat X_\mu.
\end{equation}
As is customary in the quantum chemistry community, we will use upper case letters to describe cluster operators, and lower case letters to describe cluster amplitudes, e.g., $\hat T({\bf t})$, $\hat C({\bf c})$, $\hat R({\bf r})$ or $\hat S({\bf s})$. 
By construction, acting on the reference state $\ket{\Phi_0}$ the \rev{elementary particle-hole excitation operators} define the Hilbert space of functions that are $L^2$-orthogonal to $\ket{\Phi_0}$.
We henceforth set $\VC = \{\ket{\Phi_0} \}^\perp$ and note that $\hat T({\bf t}) \ket{ \Phi_0 }\in \VC$ for all amplitudes. 
Hence, using the standard Galerkin projection approach, we can express the \rev{second equation and} the orthogonality constraint in Eq.~\eqref{eq:projections} as 
\begin{equation}
\label{eq:OrthogonalCond}
0 = \langle  \Phi_\mu  \mid \hat \ham({\bf t})  \mid \Phi_0\rangle, \quad \forall \mu
\end{equation}
where $\ket{\Phi_\mu} = \hat X_\mu \ket{\Phi_0}$ and $\hat \ham({\bf t})=e^{-\hat T({\bf t})}\hat \ham e^{\hat T({\bf t})}$. 
Since the cluster operator is defined by the cluster amplitudes $\bf t$, we can define the CC energy as a function of $\bf t$, i.e., 
\begin{equation}
    \ene_\cc({\bf t}) = \langle \Phi_0 \mid \hat \ham({\bf t}) \mid \Phi_0 \rangle .
\end{equation}
\rev{Note that the similarity transformed Hamiltonian $\mathcal{H}({\bf t})$ for all amplitudes ${\bf t}$ has the same eigenvalues as the original Hamiltonian $\mathcal{H}$. However, the CC equations in~\eqref{eq:OrthogonalCond} do not arise from diagonalizing the similarity-transformed Hamiltonian.}

We emphasize that the orthogonality conditions in Eq.~\eqref{eq:OrthogonalCond} yields a square system of polynomial equations, see e.g.~\cite{helgaker2014molecular,bartlett2007coupled,faulstich2022coupled}.
Hence, a key object in CC theory is the CC variety:   
\begin{equation}\label{eq:CC-variety}
\mathcal{S}
=
\{
{\bf t} \in \mathbb{F}^K ~|~ \langle \Phi_\mu \mid \hat \ham({\bf t}) \mid \Phi_0 \rangle = 0 \quad \forall \mu 
\} \subset \VV,
\end{equation}
where $\mathbb{F}$ is the considered number field (either $\mathbb{R}$ or $\mathbb{C}$), $K$ is the ``system size'' (given by the number of correlated electrons, the size of one-particle basis functions as well as further selection rules) and $\VV$ is the cluster amplitude space, i.e., ${\bf t} \in \VV$ if and only if $\hat T({\bf t})\ket{\Phi_0} \in \VC$. 
Note that in this work, we always assume a finite set of one-particle basis functions (i.e., orbitals), in particular, $\VV = \mathbb{F}^K$.
\rev{Note that $\mathbb{F}$ can be $\mathbb{R}$ or $\mathbb{C}$ determining if we are seeking real or complex valued amplitudes. We emphasize that although the CC polynomial coefficients are real, the roots to the polynomial system may not be. Mathematically, the more general theory describing complex valued solutions is simpler than the theory describing real valued solutions. Therefore, it is easier to consider $\mathbb{F} = \mathbb{C}$ for mathematical considerations, however, solving the truncated CC equations for $\mathbb{F} = \mathbb{C}$ may yield complex valued energies, e.g. see~\cite{faulstich2022coupled}. Moreover, the majority of quantum chemistry implementations seek real valued solutions, i.e., the CC equations are solved for $\mathbb{F} = \mathbb{R}$.}

All possible determinants that can be generated given the basis set and number of (correlated) electrons can be represented in an excitation graph, $G^\full$, where the vertices are the determinants $\ket{\Phi_\mu}$ and the edges are the operators $\hat X_\mu$~\cite{csirik2023disc}. 
However, we might not always want to consider all possible determinants that can be generated but rather a subset $G \subset G^\full$. 
For example, we might just consider excitations up to a certain order or ``rank'', such as CCSD ($\hat X_\mu$ contains excitation of at most two electrons), CCSDT ($\hat X_\mu$ contains excitations of at most three electrons), etc. Thus, we will sometimes write $\mathcal S = \mathcal{S}(G)$, $\VV = \VV(G)$, etc. to highlight that we consider a truncated CC scheme as dictated by $G\subset G^\full$. 

In the untruncated case, $G^\full$, we note that for given ${\bf t}_*\in\mathcal{S}$ the wavefunction $\Psi=(c_0 \hat I + \hat C({\bf c}))\ket{\Phi_0}$ is an eigenfunction of $\ham $, i.e., there exists a constant $\ene$ such that $\hat \ham\ket{\Psi}=\ene\ket{\Psi}$, if and only if $e^{-\hat T({\bf t}_*)}(c_0 \hat I + \hat C({\bf c}))=r_0 \hat I + \hat R({\bf r})$, where 
\begin{equation}
\left.
\begin{aligned}
\ene_\cc({\bf t_*})r_0 + \langle \Phi_0 \mid \hat \ham({\bf t}_*)  \hat R({\bf r})\mid\Phi_0\rangle  &= \ene r_0 \\
\Pi_\VC \hat \ham({\bf t}_*) \hat R({\bf r})\ket{\Phi_0} &= \ene \hat R({\bf r})\ket{\Phi_0} 
\end{aligned}
\right\}
\end{equation}
with $\Pi_\VC$ being the orthogonal projection onto $\VC$ and ${c_0 = r_0}$. 
For a proof of this statement we refer to Lemma~4.1 in~\cite{csirik2023coupled}. 

To shed some light on the above description of eigenstates using the CC framework, we will give a few examples (all familiar to the quantum-chemistry setting):

\begin{enumerate}
\item[(i)]  We first note that $\ene = \ene_\cc({\bf t}_*)$ is equivalent to ${\bf r}_0\neq 0$, $\hat R({\bf r}) = 0$, and then $\ket{\Psi}= e^{\hat T({\bf t}_*)}\ket{\Phi_0}$. 
\item[(ii)] If $\ene = \ene_\cc({\bf t}_*)$ and there is $\hat R({\bf r})\neq 0$, then $\ket{\Psi}=(r_0 \hat I + \hat R({\bf r})) e^{\hat T({\bf t}_*)}\ket{\Phi_0}$ is another eigenstate, i.e., $\ene$ is a degenerate energy level.  
\item[(iii)] If $\ene \neq \ene_\cc({\bf t}_*)$ and $\hat R({\bf r})\neq 0$, then $\ket{\Psi}=(r_0 \hat I + \hat R({\bf r})) e^{\hat T({\bf t}_*)}\ket{\Phi_0}$ is an ``excited'' eigenstate with respect to the energy level $\ene_\cc({\bf t}_*)$).
\end{enumerate}

If we let
\begin{equation}
    \ket{\Psi_{(k)}} = e^{\hat T({\bf t}_{(k)})} \ket{\Phi_0}
\end{equation}
denote the $k$th eigenstate to $\hat\ham$, the EOM-CC method~\cite{Rowe1968,monkhorst1977calculation,Koch1990,StantonBartlett1993,Koch1994} gives with $\hat R({\bf r}_{(1)})= \hat I$ that the solutions (under certain assumptions) can be written 
\begin{equation}
    \ket{\Psi_{(k)}} = \hat R({\bf r}_{(k)}) e^{\hat T({\bf t}_{(1)})} \ket{\Phi_0} .
\end{equation}
We will in the remaining part only consider the case (i) above.

The ground state (intermediately normalized) moreover corresponds to 
$\ket{\Psi_{(1)}} = e^{\hat T({\bf t}_{(1)})} \ket{\Phi_0}$, 
where
\begin{equation}
{\bf t}_{(1)} = \argmin_{{\bf t}\in \mathcal{S}} ~ \mathcal{E}_{\cc}(\bf t).
\end{equation}
We emphasize that in the untruncated case, the algebraic variety $\mathcal{S}$ (given in Eq.~\eqref{eq:CC-variety}) describes all solutions to the electronic Schr\"odinger equation, given that the eigenstates have non-zero overlap with the reference determinant $\ket{\Phi_0}$.
Hence, the $k$th eigenstate (that has non-zero overlap with the reference determinant $\ket{\Phi_0}$) can be characterized using the min-max principle, i.e., 
\begin{equation}
\ket{\Psi_{(k)}} = e^{\hat T({\bf t}_{(k)})} \ket{\Phi_0}
\end{equation}
where ${\bf t}_{(k)}$ corresponds to the argument that solves the min-max problem
\begin{equation*}
\min \lbrace \max_{{\bf t}\in \mathcal{S}_k} ~\mathcal{E}_{\cc}(\bf t) ~|~\mathcal{S}_k \subset \mathcal{S},\,|\mathcal{S}_k| = k \rbrace.
\end{equation*}

\rev{In the case of untruncated CC (FCC), amplitudes that solve the FCC equations describe an eigenstate of the Hamiltonian. Therefore, the cardinality of $\mathcal{S}$ is equal to the number of eigenstates of $\mathcal{H}$ that are intermediately normalized, hence, the number of roots is bounded by the number of Slater determiants.
However,}  when truncations are imposed, the cardinality of $\mathcal{S}= \mathcal{S}(G)$ increases, i.e., truncated CC theory yields (some) unphysical solutions. 
Therefore, it becomes less clear what the different elements in $\mathcal{S}$ describe.  
Understanding the variety $\mathcal{S}$ and characterizing (some of) its elements are the subject of this manuscript. 
To that end, we employ two homotopy continuation perspectives: 
The first is to compute $\mathcal{S}$ in its entirety, where a homotopy is used to connect $\mathcal{S}$ to solutions of a simpler system of polynomial equations. 
The second is to characterize the physical solutions in $\mathcal{S}$ corresponding to truncated CC equations, where a homotopy is used to connect $\mathcal{S}$ to the FCI solutions (or at least some ``higher'' truncation scheme, i.e., less truncated CC equations).

\section{Homotopy continuation}
\label{sec:h-cont}

Homotopy continuation methods are well studied mathematically, we refer the interested reader to~\citenum{bates2023numerical,garcia1979finding,morgan2009solving,sommese2005numerical}.
Polynomial homotopy continuation is a numerical method to compute solutions to systems of polynomial equations,
\begin{equation}
F\left(x_1, \ldots, x_n\right)=\left[\begin{array}{c}
f_1\left(x_1, \ldots, x_n\right) \\
\vdots \\
f_m\left(x_1, \ldots, x_n\right)
\end{array}\right]=0 .\end{equation}

Note that in our case, $F$ corresponds to the CC equations.
In a general case, we require $m \geq n$, however, the CC equations are a square system, i.e., $m=n$. 
The underlying idea is straightforward: to solve $F(\mathbf{x})=0$, we construct an auxiliary system of polynomial equations, called $G(\mathbf{x})=0$,  with known zeroes, together with a homotopy that connects both systems. 
More precisely, we define a family of systems $H(\mathbf{x}, \lambda)$ for $\lambda \in \mathbb{R}$ interpolating between $F$ and $G$, i.e., $H(\mathbf{x}, 0)=F(\mathbf{x})$ and $H(\mathbf{x}, 1)=G(\mathbf{x})$. 
Considering one zero, $\mathbf{y}$, of $G(\mathbf{x})$ and restricting to $\lambda \in[0,1]$, $H(\mathbf{x}, \lambda)=0$ defines a solution path $\mathbf{x}(\lambda) \subset \mathbb{C}^n$ such that $H(\mathbf{x}(\lambda), \lambda)=0$ for $\lambda \in[0,1]$ and $\mathbf{x}(1)=\mathbf{y}$. 
The path is followed from $\lambda=1$ to $\lambda=0$ to compute the solution $\mathbf{z}=\mathbf{x}(0)$. 
This is equivalent to solving the initial value problem
\begin{equation*}
\frac{\partial}{\partial \mathbf{x}} H(\mathbf{x}, \lambda)\left(\frac{\mathrm{d}}{\mathrm{d} \lambda} \mathbf{x}(\lambda)\right)+\frac{\partial}{\partial \lambda} H(\mathbf{x}, \lambda)=0, \quad \mathbf{x}(1)=\mathbf{y} .
\end{equation*}
Which is known as the Davidenko differential equation~\cite{davidenko1953new,davidenko1953approximate}. 
We say that $\mathbf{x}(1)=\mathbf{y}$ gets tracked towards $\mathbf{x}(0)$. 
For this to work, $\mathbf{x}(\lambda)$ must be a regular zero of $H(\mathbf{x}, \lambda)=0$ for every $\lambda \in(0,1]$. 
In the case of nonregular solutions at $\lambda=0$, so-called endgames are employed which are special numerical methods~\cite{morgan1992computing}.

\begin{figure}[h!]
    \centering
    \includegraphics[width = 0.45 \textwidth]{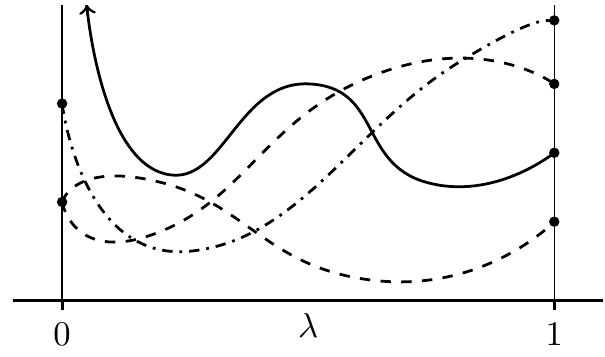}
    \caption{Sketch of possible homotopy paths. The solid line shows a path with no finite limit as $\lambda \to 0$, the dashed lines have the same limit, and the dotted-dashed line has a unique limit.}
    \label{fig:homotopies}
\end{figure}

When tracking the solutions described by the homotopy, we may encounter different scenarios, see FIG.~\ref{fig:homotopies}. 
As shown in~\citenum{bates2023numerical}, we can see that one path (solid line) has no finite limit as $\lambda \to 0$, while the other three have limits.
One path (dotted-dashed line) has a unique limit, i.e., the endpoint of one at $\lambda = 0$ is the regular zero of the target system
$F(x)$.
Two paths (dashed lines) have the same limit, and their common endpoint is an isolated zero of $F(x)$ of multiplicity two. 

\rev{We moreover wish to highlight that great progress has been made along the lines of homotopy continuation based software development exploiting parallel implementations, which can significantly extend their application areas in the coming years. In particular, we here want to highlight PHCpack~\cite{verschelde1999algorithm}, Bertini~\cite{bates2006bertini}, HOM4PS~\cite{chen2014hom4ps,lee2008hom4ps}, NAG4M2~\cite{bates2023numerical}, and HomotopyContinuation.jl~\cite{breiding2018homotopycontinuation}.}

\section{Root structure of polynomial systems}
\label{sec:RootStructure}

In this section, we will elaborate on the fundamental and practical importance of understanding the root structure of the CC equations. 
We emphasize that it is hard to describe the root structure of the CC equations in general because it is a high-dimensional and non-linear system. 
Although this seems to be a daunting task, the CC equations show a number of symmetries, which allow some results regarding the system's root structure~\cite{vzivkovic1978analytic,kowalski1998towards,piecuch2000search,kowalski2000complete2,kowalski2000complete,faulstich2018analysis}.

\begin{figure}[h!]
    \centering
    \includegraphics[width = 0.5\textwidth]{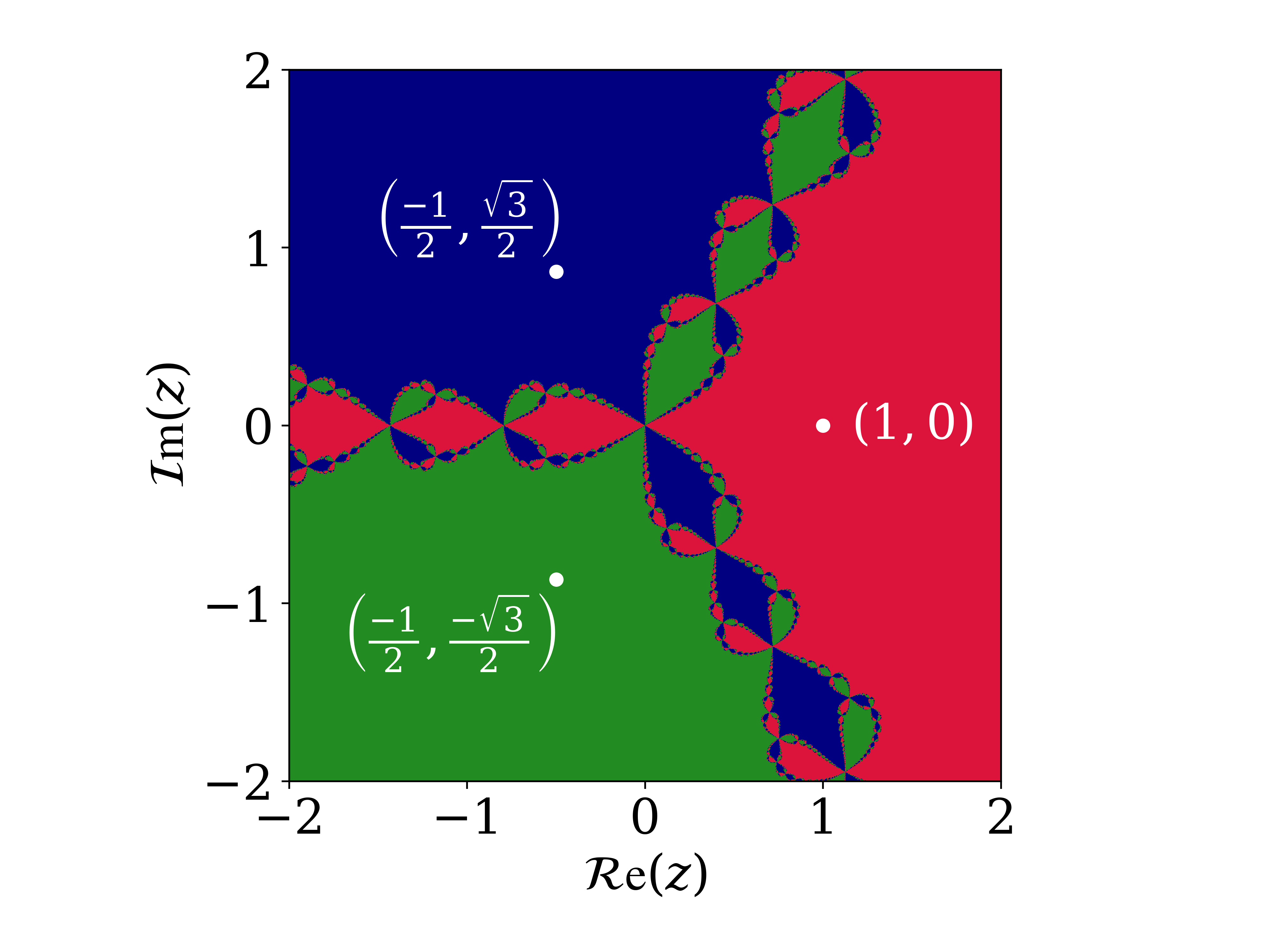}
    \caption{Newton fractal of $p(z) = z^3 -1$. The white dots correspond to the roots $x_1 = 1+i \cdot 0$ and $x_{2,3} = -1/2\pm i \sqrt{3}/2$. The different colored regions, red, blue, and green, correspond to the basins of attraction of the roots $x_1$, $x_2$, and $x_3$, respectively.}
    \label{fig:Newton_fractal}
\end{figure}

The root structure of a polynomial system is of fundamental importance as it reveals, e.g., the multiplicity of the roots or whether the roots are real or complex~\cite{hubbard2001find}. 
Having information about the root structure is also of practical importance when using approximate methods. 
Most commonly, (quasi) Newton type methods are employed to find and approximate {\it one} root of the CC equations.
However, the convergence behavior of (quasi) Newton type methods is merely locally well understood; its global convergence behavior can be highly complicated~\cite{hubbard2001find, schleicher2002number}. 
This can be illustrated through Newton fractals. 
Newton fractals are graphical representations of the iterative process used to find the roots of a given polynomial system using (quasi) Newton type methods. 
To create a Newton fractal, one assigns to each initialization the root to which it converged.  
In the case of one polynomial, one can color each point in the complex plane according to the root to which it converges under the considered (quasi) Newton type method, see e.g.~FIG.~\ref{fig:Newton_fractal} where we show the Newton fractal for $p(z) = z^3-1$.

Clearly, FIG.~\ref{fig:Newton_fractal} is a simplified perspective, since it shows the root structure of merely one polynomial.  However, it already illustrates the (potential) issues that the application of (quasi) Newton type methods to the CC equations can have. 
Figure~\ref{fig:Newton_fractal} shows the three roots of the polynomial $p(z) = z^3-1$, namely $x_1 = 1+i \cdot 0$ and $x_{2,3} = -1/2\pm i \sqrt{3}/2$, as well as the color-coded basins of attraction that correspond to the respective roots. 
We can clearly see that if the initial guess is {\it close} to one of the solutions, (quasi) Newton type methods will stably converge to that solution. 
However, if the initial guess is not close to a solution, but lies in one of the ``fractal branches'', the convergence of (quasi) Newton type methods becomes very unstable~\cite{eriksen2019many}. 
In fact, the numerical stability can become so poor that differences at the level of machine precision will change the converged result. 
Since the CC equations are so high-dimensional, a bruit force visualization is already for the smallest systems impractical. 
Yet, since the highly intricate global convergence behavior of (quasi) Newton type methods is rather common, one must be aware of this phenomenon, especially in a scenario when perturbative initializations are not justifiable. 
In fact, if the initial guess lies in a ``fractal branch'', the employed (quasi) Newton type methods may yield a root that is suboptimal, although the used CC ansatz (e.g.~CCSD) is well capable of describing the targeted state.  
To summarize this section, using (quasi) Newton type methods to approximate a root of the CC equations that corresponds to a targeted state, one must ensure to be in the correct basin of convergence. Otherwise, it is unclear if the obtained CC result is an approximation to the targeted state.   

\section{Homotopy I}
\label{sec:Homotopy1}

In this section, we discuss a path toward numerical methods that aim to find all solutions to the CC equations. 
Due to the complicated structure of the basins of attraction for (quasi) Newton type methods described in the previous section, employing this approach for finding a CC solution heavily relies on certain local assumptions of the initial guess, which mathematically reduces to a perturbative picture. 
\rev{This importance of perturbation theory related to CC theory is well-known in the computational chemistry community, and we refer the interested reader to~\cite{bartlett2007coupled} for a thorough and text-book-like presentation. 
However, this perturbative argument is also well-known to fail. 
In particular, we wish to highlight important failure modes presented in~\cite{piecuch1990coupled} revealing the existence of algebraic branch points as a perturbation is turned on. 
The existence of such branch points strongly indicates that the commonly employed procedure of approximating {\it one} root via (quasi) Newton-type methods starting from the Hartree-Fock reference cannot be generally applied.
}
Therefore, in order to reliably extend the CC theory to the non-perturbative regime, accessing all roots of the CC equations appears to be inevitable.
Here, homotopy continuation methods are one potential way to achieve this, but the high dimensionality of the equations makes a direct application of these types of procedures challenging~\cite{faulstich2022coupled,piecuch2000search}.

In fact, establishing a good enough bound to the number of roots is already challenging! 
This bound is important since it is used to initialize an auxiliary system $G$, the employed homotopy continuation method will then track all solutions of $G$. 
Clearly, when over-estimating the number of roots, along the path $\lambda \to 0$, spurious roots will collapse, yet, if the estimated number of roots is too large the produce will simply become numerically intractable. 
A trivial bound can be obtained by observing that Hadamard's lemma, i.e., the expansion of $\ham({\bf t})$, together with commutator considerations (see e.g.,~\cite{helgaker2014molecular,bartlett2007coupled}) yields a polynomial system order of less or equal than four.
The corresponding B\'ezout bound is then 
\begin{equation}
\label{eq:bezout_0}
\mathcal{N} \leq 4^{n_{\mathcal{K}}},
\end{equation}
where $\mathcal{K}$ describes the number of projective equations. 
However, this number grossly overestimates the number of roots, since no further structure of the CC equations is taken into account.  
A first dramatic reduction can be obtained by noticing that the projective equations onto the singly excited Slater determinants are of the order less or equal to three (see e.g.~\cite{helgaker2014molecular,bartlett2007coupled}). 
Since our main objective is to incorporate homotopy methods for CCSD, we will \rev{for the remainder of this section only consider this truncation level}.
The above consideration then yields
\begin{equation}
\label{eq:bezout_1}
\mathcal{N} \leq 3^{n_s} 4^{n_d}
\end{equation}
where $n_s$ and $n_d$ are the number of projected single and double equations, respectively~\cite{piecuch2000search}. 
\rev{Note that Eq.~\eqref{eq:bezout_1} is specific to the CCSD scheme, and generalizations to this bound for higher-order CC schemes, such as CCSDT and CCSDTQ are also discussed in~\cite{piecuch2000search}.}

Although this bound includes the further structure of the CCSD equations, it is still overestimating the true number of roots. 
As has been recently shown, this bound can be significantly improved by rewriting the CCSD equations as a quadratic system~\cite{faulstich2022coupled}.
The key in this reformulation is to notice that the anti-symmetry property of Slater determinants allows for factorizing disconnected doubles. 
To that end, we define the variety $A\subseteq  \mathbb{F}^{n_s+ 2n_d}$ (where $\mathbb{F}$ is either $\mathbb{R}$ or $\mathbb{C}$) over which we seek to optimize the CCSD equations.

Formally, we define an index map $\iota$ that flattens the tuple $(i,j,a,b)$ and off-sets this compound index by $n_s+ n_d$. 
Note that since $i$, $j$ are occupied indices, and $a$, $b$ are virtual indices, there are exactly $n_d$ auxiliary indices that are obtained by $\iota$. 
We then define the variety $\tilde{ \mathcal{S}}$ as
\begin{equation*}
\tilde{\mathcal{S}} = \{x\in \mathbb F^{n}| x_k - x_i^ax_j^b+x_i^bx_j^a= 0,\, \forall k = \iota (i,j,a,b)\},
\end{equation*}
where we introduced the short-hand notation ${n = n_s + 2n_d}$.
On this variety, every projected equation $f_\mu$ takes the form
\begin{equation}
\label{eq:OrderRed}
f_\mu (\Pi_{\mathbb{V}} {\bf x})= \sum_{q,r} h^\mu_{q,r}x_qx_r
\end{equation}
where $h^\mu_{q,r}$ is a matrix (see~\cite{faulstich2022coupled}), $\Pi_{\mathbb{V}} $ is the projection onto $\mathbb{V}$ and $\bf x \in \tilde{\mathcal{S}}$.
Note that the l.h.s.~in Eq.~\eqref{eq:OrderRed} is a potentially fourth-order polynomial in $n$ variables, whereas the r.h.s. in Eq.~\eqref{eq:OrderRed} is a second order polynomial in $n+n_d$ variables.  
This order reduction yields the improved bound of 
\begin{equation}
\mathcal{N} \leq 2^{n_s+2n_d},
\end{equation}
which is exponentially better than the existing B\'ezout bound~\eqref{eq:bezout_1}.

Although this reduction is already quite significant, we expect that for special cases further symmetry considerations can be incorporated yielding even better bounds. 
However, B\'ezout type bounds are always worst-case estimates. 
A better bound is expected when employing the Bernstein-Khovanskii-Kushnirenko (BKK) theorem. 
B\'ezout type bounds arise from the idea that the polynomials in the considered system are independent of each other, in which case the number of roots corresponds to the product of the individual number of roots; hence, they are worst-case bounds. 
BKK-type bounds are less intuitive since the BKK theorem relates the root counting problem for a system of polynomial equations with the theory of convex bodies. 
More precisely, the BKK theorem shows that the generic number of isolated solutions to a system of (Laurent) polynomial equations equals the mixed volume of the Newton polytopes of the (Laurent) polynomials.
The general challenge when establishing a BKK-type bound is to compute the volume of convex bodies, i.e., the Newton polytopes. 
However, general algorithms for computing the mixed volume are exponential in the dimension. 
Since even computing the volume is known to be \#P-hard, a brute force approach seems doomed. 
That being said, the CC equations are very structured, and a surrogate system of Newton polytopes can be established~\cite{faulstich2022coupled}.

\section{Homotopy II}
\label{sec:Homotopy2}
As already mentioned above, 
the second form of homotopy approaches we wish to discuss is the one used to characterize the physical solutions in $\mathcal{S} = \mathcal{S}(G)$ corresponding to truncated CC equations (as characterized by the choice of excitation graph $G \subset G^\full$). A homotopy scheme can then be used to connect $\mathcal{S}(G)$ to some ``higher'' truncation scheme, in particular the FCC (FCI) regime (i.e., CC method with $G^\full$). 

Before continuing further, we note that ${\bf t} \in \mathcal{S}(G)$ is equivalent to 
\begin{equation}
    \langle\hat S({\bf s}) \Phi_0 \mid \hat \ham({\bf t}) \mid \Phi_0\rangle = 0 
\end{equation}
for all ${\bf s} \in \VV(G)$. Let $\VV^*$ denote the dual space of $\VV$, we then define the CC mapping $\opA({\bf t}): \VV \to \VV^*$ via  
\begin{equation}
    \langle {\bf s},\opA({\bf t}) \rangle = \langle\hat S({\bf s}) \Phi_0 \mid \hat \ham({\bf t}) \mid \Phi_0\rangle.
\end{equation}
Here $\langle \cdot,\cdot\rangle$ is also used to denote the dual pairing between amplitudes and elements in the dual amplitude space. 
Note that, by definition, $\mathcal S$ is the set of zeros to $\opA$. We will assume for ${\bf t}_* \in \mathcal S$ 
that $\det \opA'({\bf t}_*)\neq 0$, i.e., there are no non-trivial right eigenvectors to $\hat \ham ({\bf t}_*)$ corresponding to the 
eigenvalue $\ene_\cc({\bf t}_*)$ (in short, ${\bf t}_*$ is a non-degenerate zero).

Let $\VV^1$ be the amplitude space corresponding to the excitation graph $G^1$. This can be thought of as the untruncated CC case, i.e., full graph with all possible determinants (vertices) included. Similarly, $\VV^0$ is the amplitude space corresponding to $G^0$, which contains amplitudes with rank $\le\rho$ ($2\leq \rho < N$). We then decompose $\VV^1=\VV(G^1)$ as: $\VV^1=\VV^0\oplus\VV^\perp$. 
Here, $\VV^\perp$ contains amplitudes with rank rank strictly greater than $\rho$. 
It hols $\langle {\bf s}^0, {\bf s}^\perp\rangle = 0 $ for ${\bf s}^0 \in \VV^0$ and ${\bf s}^\perp\in \VV^\perp$.

\subsection{Existence for Kowalski--Piecuch homotopy}

We will next discuss (from a more mathematical perspective) how to make sense of a truncated CC solution by finding a trajectory that connects it to a FCC (FCI) solution.
We define the {\it Kowalski--Piecuch homotopy} $\opK_{\KP} : \VV^1\times[0,1]\to(\VV^1)^*$ via the instruction
\begin{equation}
    \begin{aligned}
        &\dua{\opK_{\KP}({\bf t}^1,\lambda)}{{\bf s}^1}=
        \langle \hat S({\bf s}^0)\Phi_0 \mid \hat \ham({\bf t}^0) \mid \Phi_0\rangle \\
        &\quad + \langle \hat S({\bf s^\perp})\Phi_0 \mid \hat \ham({\bf t}^1) \mid \Phi_0\rangle \\
 &\quad + \lambda \langle \hat S({\bf s}^0) \Phi_0 \mid \hat \ham({\bf t}^0) \mid ( e^{\hat T({\bf t}^\perp)} - \hat I ) \Phi_0 \rangle 
    \end{aligned}
\end{equation}
for all ${\bf t}^1,{\bf s}^1\in \VV^1$ and $\lambda\in[0,1]$.
We can note that by construction, the Kowalski--Piecuch homotopy satisfies $\opK_{\KP}({\bf t}^1,1)=\opA({\bf t}^1)$. 
Furthermore, $\opK_{\KP}({\bf t}^1_{**},0)=0$ is equivalent to   
\begin{align}\label{eq:usual-CC-tr}
\langle \hat S({\bf s}^0)\Phi_0 \mid \hat \ham({\bf t}_{**}^0) \mid \Phi_0\rangle &=0, \\
\langle \hat S({\bf s}^\perp) \Phi_0 \mid \hat \ham({\bf t}_{**}^0+{\bf t}_{**}^\perp) \mid \Phi_0\rangle &=0. \label{eq:KP-aux}
\end{align}
Equation~\eqref{eq:usual-CC-tr} is the usual truncated CC equation (associated with $G^0$) and Eq.~\eqref{eq:KP-aux} is the Kowalski--Piecuch-auxiliary equation. 

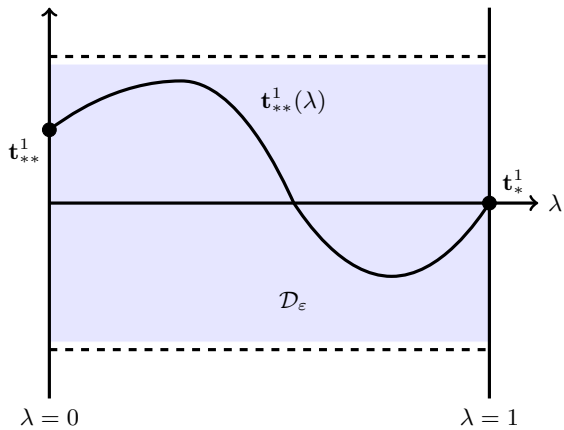
\begin{figure}
\begin{tikzpicture}[scale=.65] 

\node[rectangle,
    draw = white,
    fill = blue!10!white,
    minimum width = 5.9cm, 
    minimum height = 3.7cm] (r) at (4.5,7) {}; 

\draw[->,very thick] (0,3)  node[below] {$\lambda=0$} -- (0,11);
\draw[very thick] (9,3) node[below] {$\lambda=1$} -- (9,11);
\draw[->,very thick] (0,7) -- (10,7) node[right]{$\lambda$};

\draw[dashed,very thick] (0,10) -- (9,10);
\draw[dashed,very thick] (0,4) -- (9,4);

\draw[very thick] (0,8.5) parabola bend (2.7,9.5) (5,7);
\draw[very thick] (5,7) parabola bend (7,5.5)  (9,7);

\filldraw[black] (9,7) circle (4pt) node[above] {$\qquad {\bf t}_*^1 $};
\filldraw[black] (0,8.5) circle (4pt) node[below] {${\bf t}_{**}^1 \qquad {}$};

\fill (5,5) node{$\mathcal D_\varepsilon$};
\fill (5,9.1) node{${\bf t}_{**}^1(\lambda)$};

\end{tikzpicture}
\caption{Visualization of the existence result of the Kowalski--Piecuch homotopy. Under certain assumptions there is a ``tube'' in amplitude space that connects a truncated CC solution to a FCC (FCI) solution. In principle, such trajectory allows us to select which solutions of a truncated CC calculation are ``physical''. }
    \label{fig:ExiKP}
\end{figure}

The Kowalski--Piecuch homotopy has been used to establish an existence result that connects a truncated solution to a corresponding untruncated (``full'') solution~\cite{csirik2023coupled}: 

Let ${\bf t}_*^1\in\VV^1$ be a non-degenerate zero of $\opA$. 
Under technical assumptions (see Theorem~4.34 in~\cite{csirik2023coupled}), we can find $\varepsilon>0$ such that for any $\lambda\in[0,1)$ there exists ${\bf t}_{**}^1(\lambda)\in \mathcal D_\varepsilon$ fulfilling $\opK_\KP({\bf t}_{**}^1(\lambda),\lambda)=0$, where 
$$
\mathcal D_\varepsilon=\{ {\bf t}_*^1 + {\bf r}^1 \in \VV^1 : \|{\bf r}^0\|_\VV^2 + \|{\bf r}^\perp\|_\VV^2 < \varepsilon  \}.
$$
In particular, there exists ${\bf t}_{**}^1\in \mathcal D_\varepsilon$ such that $\opA({\bf t}_{**}^0)=0$.
(See FIG.~\ref{fig:ExiKP} for an illustration.)

For the interested reader, the technical assumptions are discussed in~\cite{csirik2023coupled} (in Theorem 4.34 and directly afterward). 
In essence, these assumptions concern the fluctuation potential (and how it couples the truncated space to the ``rest'', which can be controlled by the truncation level $\rho$) and the size of the ${\bf t_{**}}^\perp$ (that cannot be too large). 
One of the assumptions can also be interpreted as a perturbative assumption related to the size of the second derivative of the CC mapping (i.e., $\opA''$) at ${\bf t}_*^1$.  

\subsection{Error estimate for Kowalski--Piecuch homotopy}

The {\it Fundamental theorem of the formalism of $\beta$-nested equations} of Kowalski and Piecuch~\cite{piecuch2000search} can be cast into an {\it a posteriori} error estimate~\cite{csirik2023coupled}. 
Let $\VV^1=\VV(G^\full)$. Suppose that 
${\bf t}_*^1\in\VV^1$ is a zero of $\opA$ and ${\bf t}_{**}^1\in\VV^1$ a zero of $\opK_\KP(\cdot,0)=0$. We then have the energy error estimate valid for any (approximate) eigenstate (and not just the ground state) parameterized by the truncated amplitudes ${\bf t}_{**}^0$ that solve Eq.~\eqref{eq:usual-CC-tr}:

Let $\rho\ge 2$ and assume that the nonorthogonality condition $\langle e^{\hat T({\bf t}^0_{**})} \Phi_0 \mid e^{\hat T({\bf t}_*^1)}\Phi_0\rangle\neq 0$ hold.
Then
\begin{equation}\label{eq:E-error}
|\ene_{\cc}({\bf t}_{**}^0)-\ene_{\cc}({\bf t}_*^1)|\le C({\bf t}_{**}^1,{\bf t}_{*}^1) \,\,\|{\bf t}_{**}^\perp\|_\VV ,
\end{equation}
where ${\bf t}_{**}^\perp$ solves Kowalski--Piecuch-auxiliary equation~\eqref{eq:KP-aux}.

Here the constant in the rhs. of Eq.~\eqref{eq:E-error} is given by 
\begin{equation}
\begin{aligned}
    C({\bf t}_{**}^1,{\bf t}_{*}^1)
      = \tilde C({\bf t}_{**}^1,{\bf t}_{*}^1) \,
     |\langle e^{\hat T({\bf t}^0_{**})} \Phi_0 \mid e^{\hat T({\bf t}_*^1)}\Phi_0\rangle|^{-1} .
\end{aligned}
\end{equation}
\rev{
The constant $\tilde C$ is a multifaceted quantity that comprises several contributing factors. These factors include a norm equivalence constant, the similarity transform of the fluctuation operator (where the Hamiltonian is a sum of the Fock and fluctuation operator), and the size of the orthogonal space $\VC^\perp$. 
We emphasize that the constant becomes smaller in the perturbative regime of the CC method when the fluctuation potential is small and $\VC^\perp$ is less physically relevant. 
For more details, we refer the interested reader to Theorem~4.40 in~\cite{csirik2023coupled}.}

If the nonorthogonality condition holds and ${\bf t}_{**}^\perp=0$, then ${\bf t}_{**}^0$ is a FCC (FCI) solution and the energy error estimate implies that the error is zero, i.e., $\ene_\cc({\bf t}_{**}^0)=\ene_{\cc}({\bf t}_*^1)$. 
Consequently, $\|{\bf t}_{**}^\perp\|_\VV$ allows us to view Eq.\eqref{eq:KP-aux} as providing an \emph{a posteriori} error estimate for a truncated CC calculation (that gives us ${\bf t}_{**}^0$), which could then have practical use if the Kowalski--Piecuch-auxiliary equation can be solved (at least approximately due to its FCI complexity). 

If $\langle e^{\hat T({\bf t}^0_{**})}  \Phi_0 \mid e^{\hat T({\bf t}_*^1)}\Phi_0 \rangle=0$ and  ${\bf t}_*^1$ is assumed to be non-degenerate, then 
$e^{\hat T( {\bf t}^0_{**})} \ket{\Phi_0}$ and $e^{\hat T({\bf t}_*^1)}\ket{\Phi_0}$ represent different eigenstates. 
This means that $e^{\hat T({\bf t}^0_{**})} \ket{\Phi_0}$ is an approximation to an eigenstate \emph{different} from $e^{\hat T({\bf t}_*^1)}\ket{\Phi_0}$. 
In this case, it does not make sense to try to connect these solutions and a comparison between the energies in terms of an energy error estimate is not meaningful (in agreement with a divergent constant $C$). 
Conversely, if the energy difference $|\ene_{\cc}({\bf t}_{**}^0)-\ene_{\cc}({\bf t}_*^1)|$ diverges for finite cluster amplitudes, then $\langle e^{\hat T({\bf t}^0_{**})}  \Phi_0 \mid e^{\hat T({\bf t}_*^1)}\Phi_0 \rangle\to 0$ since we 
are trying to connect solutions at different truncation levels that are orthogonal (without degeneracies this means solutions belonging to different eigenvalues).

\rev{In the beginning of this section we referenced the origin of the energy error estimate above to the 
the Fundamental theorem of the formalism of $\beta$-nested equation due to Kowalski and Piecuch~\cite{piecuch2000search}.
This theorem offers an explicit formula for the noniterative correction required to supplement the energy obtained from an (approximate) SRCC calculations, e.g., on the CCSD level, in order to retrieve the FCI result (see Eq.~(6) in~\cite{KowalskiPiecuchMMCC-2000}). The approach of addressing the many-electron correlation problem, based on this formula, is called the \emph{method of moments of coupled-cluster equations} (MMCC)~\cite{piecuch2000search,KowalskiPiecuchMMCC-2000} 
In this regard the Kowalski--Piecuch homotopy has also helped the development of the completely renormalized (CR) CC and EOM-CC methods 
and generalizations that exist under the name CC$(P;Q)$ (see~\cite{ShenPiecuchCCPQ-2012} and references therein).  
}

\section{Discussion}\label{sec:conclusion}

In this manuscript, we elaborated on recent advances in applying homotopy methods to CC theory.
We introduced the concept of homotopy continuation methods, which have proven successful in solving polynomial systems in the wider applied mathematics community. 
We then delved into the challenges faced by state-of-the-art approximation techniques when solving the CC equations in a non-perturbative regime, and we discussed how homotopy continuation methods can potentially overcome these obstacles.

Despite the promising benefits and previous research by pioneers such as Živković and Monkhorst~\cite{vzivkovic1978analytic}, as well as Kowalski and Piecuch~\cite{kowalski1998towards,piecuch2000search,kowalski2000complete2,kowalski2000complete}, the high dimensionality, non-linearity, and steep scaling of algebro-computational methods have hindered widespread adoption of homotopy continuation methods in CC theory. However, significant progress has been made in scaling algebro-computational methods~\cite{verschelde1999algorithm,bates2006bertini,chen2014hom4ps,lee2008hom4ps,breiding2018homotopycontinuation}. We are convinced that in the near future, these approaches will be adopted and further developed to be applied to CC theory. In this manuscript, we present two exciting use cases where homotopy methods can make a significant impact.

The first application involves using homotopy methods to compute all roots of the CC equations~\cite{faulstich2022coupled}. 
We outline the advantages of having access to all roots in CC theory, and how this can extend CC methods beyond an application in the perturbative regime. 
The main challenge is that, despite the great progress in the field, the scaling of algebro-computational methods is still unfavorable for high-dimensional problems like the CC equations.  
However, computational algebraic tools are developed to find the roots of polynomial systems in the most general sense, in particular, they do not exploit the structured nature and symmetries of the CC equations. 
These provide opportunities to overcome unfavorable scaling, which is the subject of current investigations.
In this article, we outlined the first step for homotopy approaches, which is an accurate estimate of the number of roots. 
We here build upon results by Kowalski and Piecuch~\cite{piecuch2000search} that reduced the estimated number of roots to $3^{n_s} 4^{n_d}$. 
We elaborate on how to reduce the order of the CC polynomials to be of merely second order by extending the CC equations to an algebraic variety $\tilde{\mathcal{S}}$. 
The obtained bound is $2^{n_s+2n_d}$, which provides a significant improvement to the existing bound.

The second use case involves homotopy methods to continue CC solutions from one level of accuracy to another, in particular, connecting solutions from a truncated CC scheme to the FCI regime. This specific use of homotopy is, in principle, a very important tool to justify what elements of $ \mathcal{S}(G)$ constitute physical solutions that approximate eigenstates. The Kowalski--Piecuch homotopy (with parameter $\lambda\in [0,1]$) allows us to create trajectories that connect solutions at different levels of accuracy and we have in this article discussed a mathematical existence result that guarantees such curves as a function of $\lambda$ (under very technical assumptions that we have here omitted). Furthermore, building on the $\beta$-nested equations of Kowalski and Piecuch, an {\it a posteriori} 
energy error estimate has previously been derived. This result is not restricted to the ground state. 
It would be interesting if future work could reveal more about the role of the Kowalski--Piecuch-auxiliary equations that might be of practical use. 
It is noteworthy how these auxiliary equations bear a resemblance to the tailored CC method~\cite{kinoshita2005coupled,hino2006tailored} that might provide further insights.

\section{Acknowledgment} 
AL acknowledges funding through ERC StG REGAL under agreement No.\ 101041487 and RCN (the Research Council of Norway) under agreement 287906 (CCerror) and 262695 (Hylleraas Centre for Quantum Molecular Sciences). 
F.M.F acknowledges the funding from the Air Force Office of Scientific Research under award number FA9550-18-1-0095 and from the Simons Targeted Grants in Mathematics and Physical Sciences on Moir\'e Materials Magic.

\bibliography{lib.bib}

\end{document}